\documentclass[a4paper]{revtex4}
\usepackage[english]{babel}
\usepackage{array,booktabs}% http://ctan.org/pkg/{array,booktabs}
\usepackage{array} 
\usepackage{lipsum}   
\usepackage{calc}
\usepackage{pdflscape}
\usepackage{color}
\usepackage{float}
\usepackage[font=small,labelfont=bf]{caption}
\usepackage{graphicx}
\usepackage{amsmath}
\usepackage[nodisplayskipstretch]{setspace}

\usepackage{setspace}
\usepackage{tabularx}

\usepackage{float}
\usepackage{color}
\usepackage{amsmath}
\usepackage{float}
\usepackage{calc}
\usepackage{pdflscape}
\usepackage{color}
\usepackage{float}
\usepackage{subfigure}
\usepackage[font=small,labelfont=bf]{caption}
\usepackage{graphicx}
\begin{document}{\setlength\abovedisplayskip{4pt}}

\title{Analytical Soft SUSY Spectrum in Supersymmetric Models in Light of $ S_{4} \times Z_{n} $ flavor symmetric SUSY SO(10) theory.}

\author{Gayatri Ghosh}
\email{gayatrighsh@gmail.com}
\affiliation{Department of Physics, Gauhati University, Jalukbari, Assam-781015, India}

%\ead{{\color{blue}kalpana@gauhati.ac.in}}
%\ead{{\color{blue}gayatrighsh@gmail.com}}
%\ead{{e-mail: kalpana.bora@gmail.com}}

\begin{abstract}
The heavy right-handed neutrinos in supersymmetric models can act as the source of lepton flavor violation (LFV). LFV processes like $ \mu  \rightarrow e  \gamma $, $ \tau \rightarrow \mu \gamma $, $ \tau  \rightarrow e  \gamma $ is an effective way to explore new physics beyond the SM. Among the possible processes, $ \mu $ decays have the
greatest discovery potential in most of the supersymmetric models. Experimental inference of lepton flavor-violating processes within a supersymmetric type-II seesaw framework in the non-universal Higgs model (NUHM) and non-universal Scalar Mass model for Yukawa mixing scenarios in the $ S_{4} $ theory with an additional discrete symmetry is presented. The numerical analysis includes full 2 loop
renormalization group running effects for the the above mentioned Yukawa coupling matrices. The projected discovery reach of LFV experiments (MEG-II) is mentioned and  
those regions in mSUGRA, NUHM, NUSM models that have already been excluded by the LHC searches or that which is probed by MEG experiments is specified here. The results presented in this work can influence experimental challenges and physics motivations to construct various BSM theories and sensitivity to test these theories at next run of HE/HL LHC is also considered.
\end{abstract}
\maketitle
\section{Introduction}
\label{intro}
Supersymmetry (SUSY) remains a potent and powerful model for physics Beyond the Standard Model, despite the fact that there is insufficiency of signals from the Large Hadron Collider (LHC) {\color{blue}\cite{1}}. SUSY solves the naturalness problem i.e, the big hierarchy problem {\color{blue}\cite{2}}, introduces a stable dark matter candidate, neutralino {\color{blue}\cite{3}} and explains neutrino masses when supplemented with right-handed (RH) Majorana neutrinos {\color{blue}\cite{4}}. Hitherto, the Minimal Supersymmetric Standard Model (MSSM) {\color{blue}\cite{5}} has been confronted and yet it has sustained itself into following experimental tests: 1. Gauge coupling unification of three fundamental forces {\color{blue}\cite{6}}, 2. Discovery of the top quark within 100-200 GeV for a successful radiative electroweak symmetry breaking (REWSB) {\color{blue}\cite{7}}, 3. discovery of the Higgs boson $m_{h} \sim $ 125 GeV, within the narrow range of MSSM allowed values {\color{blue}\cite{8}}.
\par 
Neutrino oscillations, proved by experiments, requires one to go to physics beyond standard model. These neutrino oscillations, and hence mixings, are also surmised to induce lepton flavor violations in the charged leptonic sector. Theoretically, such cLFV processes could be instigated in different theories with BSM particles such as SUSY GUT  {\color{blue}\cite{Sidori2007}}, SUSY See Saw {\color{blue}\cite{S.Antusch2006}}-{\color{blue}\cite{A.Masiero2004}} , LHC Higgs Model {\color{blue}\cite{M.Blank2010}} and models with extra dimension {\color{blue}\cite{K.Agashe2006}}. In this work cLFV decay $ \mu \rightarrow e  \gamma $, getting contributions from neutrino oscillations and mixings is considered.
\par 
The current sensitivity of precision charged lepton flavor violation experiments, cLFV measurements already explores certain regions of SUSY parameter space, mostly the constrained MSSM (cMSSM or mSUGRA) {\color{blue}\cite{16}} with light scalar masses and right-handed neutrinos (RHN). The MEG Collaboration announced an upper bound for the branching fraction of the process $ \mu \rightarrow e  \gamma $: BR$(\mu \rightarrow e  \gamma )< 4.2 \times 10^{-13} $ at 90 percent C.L. {\color{blue}\cite{17}}. The expected
sensitivity of the MEG-II experiment is 6 $ \times 10^{-14}$ for three years of data taking {\color{blue}\cite{44}} .
\par 
Nevertheless precision experiments cannot be speculated of as a replacement for LHC, they can be compatible. Constructive results for sparticle searches at the LHC in company with LFV search results would constrain many BSM theories that are presently consistent with observations. 
\par 
Numerous processes implicating lepton flavor violating decays could be feasible such as $\mu \rightarrow e $, $ \tau \rightarrow \mu  $ or $ \tau  \rightarrow e $ transitions.  Enhancement for $ \mu  \rightarrow e  \gamma $ decay at next phase of MEG experiment is expected to reach BR ( $ \mu  \rightarrow e + \gamma $) $ \leq $   6$\times 10^{-14}$ {\color{blue}\cite{Adam}},{\color{blue}\cite{Baldini}}. In this work, the decay $ \mu  \rightarrow e + \gamma $ is considered, as this is top foremostly constrained by experiments. Such experimental hunt, and abstract studies on lepton flavor violation (LFV) can help us compel the new physics or BSM theories, that could be contemporary just above the electroweak scale, or within the jurisdiction of next run of LHC.  It is significant, that in the next run of LHC, i.e High Luminosity LHC (HE/HL-LHC), the center of mass energies are conventional to go to 27 TeV {\color{blue}\cite{Christoph}}. The flavor physics programme at the HL-LHC comprises many different probes- the weak decays of $ \mu $, $ \tau $ leptons and the Higgs boson in which the experiments can search for signs of new physics. It will be exciting to see the full potential of the HE-LHC to serve as a facility for precision new physics for decades to come.
\par
It is needless to say that SUSY GUTs gives intensification to tiny neutrino masses via see saw mechanisms in which noteworthy benefaction to lepton flavor violation  could come from flavor violations amidst heavy sleptons. The outcome of lepton flavor violation could become outstanding due to radiative corrections to Dirac Neutrino Yukawa Couplings (DNY), which might become apparent if the see saw scale is slightly lower than the GUT scale {\color{blue}\cite{L.Cabbibi2012}},  {\color{blue}\cite{I.H.Lee}}-{\color{blue}\cite{E. Arganda}}. Alike studies in contrastive see saw mechanisms have been carried out in {\color{blue}\cite{L.Cabbibi2012}},{\color{blue}\cite{I.H.Lee}}-{\color{blue}\cite{Hambye}}. 
In {\color{blue}\cite{L.Cabbibi2012}}, similar studies were done in scenario when neutrino masses and mixings appear attributable to type I See Saw mechanism of SUSY SO(10) theory, where the Dirac neutrino Yukawa
 couplings were of the kind- $ Y_{\nu} = Y_{u} $ and $ Y_{\nu} = Y_{u}^{diag}
 U_{PMNS}$, where $ Y_{u} = V_{CKM}Y_{u}^{diag}V_{CKM}^{\dagger}$. In this work Similar studies are done in type II See Saw scenario in $ S_{4} \times Z_{n} $ flavor symmetric model. Lepton Flavor Violation in SUSY type II seesaw {\color{blue}\cite{J.Schechter}} models have also been considered untimely in {\color{blue}\cite{Rossi}}-{\color{blue}\cite{E. Arganda}}.
\par
In this work examination on LFV decay ($ \mu  \rightarrow e  \gamma $)  using type II see saw mechanism in $ S_{4} \times Z_{n} $ flavor symmetric SUSY SO(10) theories {\color{blue}\cite{bd}} is carried out, and hence the reactivity to try out the surveillance of sparticles at HE/HL run of LHC {\color{blue}\cite{Christoph}}, in mSUGRA, NUHM,and NUSM {\color{blue}\cite{utpal}} models is also discussed. Such studies in Non Universal Gaugino Mass models were done earlier in {\color{blue}\cite{Profumo,Ga}}. It is noteworthy that $ S_{4} \times Z_{n} $ flavor symmetric SUSY SO(10) theory gives correct fit to observed neutrino oscillations and mixings and provides specific mass
textures for the quarks and leptons with only a small number of parameters and also predicts quark lepton mass relations and mixing angles in both the quark and the lepton sector. Specially, the model leads to tri-bi-maximal form for the PMNS matrix in the leading order with corrections to this imminent from charged lepton fields. In {\color{blue}\cite{L.Cabbibi2012,Ga}}, similar studies were done employing type I see saw formula, applying older value of BR($ \mu \rightarrow e \gamma $) {\color{blue}\cite{MEGJ.Adam}}. The structure of Dirac neutrino Yukawa couplings from {\color{blue}\cite{bd}} is used here in this work, for $ \tan\beta = 10-60 $, and $M_{GUT}= 2\times 10^{16}$ GeV. The value of Higgs mass as measured at LHC {\color{blue}\cite{Christoph}} and improved precision values of reactor mixing angle $ \theta_{13} $ as measured at Daya Bay, Reno {\color{blue}\cite{tor}} have been utilised in this work. Few studies on LFV in SO(10) GUTs have also been conferred in {\color{blue}\cite{Tatsuru}},{\color{blue}\cite{Amon}}. 
\par 
The minimal supergravity model (mSUGRA) is a  well driven and induced model {\color{blue}\cite{Chamseddine}}, in which Minimal Supersymmetric Standard Model (MSSM) {\color{blue}\cite{Tata}} can be inserted. In mSUGRA, SUSY is broken in the hidden sector, and is proclaimed to the visible sector MSSM fields by dint of gravitational interactions. Formation of gaugino masses {\color{blue}\cite{cremmer}} in mSUGRA (N=1 supergravity) embraces two scales $-$ spontaneous SUGRA breaking scale in the hidden sector over the singlet chiral superfield and the another one is GUT breaking scale via the non singlet chiral superfield {\color{blue}\cite{Chamseddine}}. In postulate these two scales can be distinct. But in a minimalistic perspective, they are usually assumed to be similar {\color{blue}\cite{Chamseddine}}. This conducts to a common mass $m_{0}$ for all the scalars, a common mass $M_{1/2}$ for all the gauginos and a common trilinear SUSY breaking term $A_{0}$ at the GUT scale, $M_{GUT}\simeq2\times10^{16}$ GeV. 
\par   
Now, the universality of sfermion masses, assumed in mSUGRA, NUHM models is analysed here. SO(10) symmetric soft terms approximately mean boundary conditions close to NUHM. In the scheme of SO(10) theories, all the matter fields, and the right handed neutrino, are there in the same 16-dimensional representation, and consequently, all the matter fields will have identical mass at the high scale. Nevertheless, the higgs fields can have dissimilar mass, as they doesnot survive in the same representation as the matter field. Therefore, the boundary terms for the SO(10) theory are compatible with NUHM and mSUGRA (in mSUGRA, all the higgs will be in equivalent representation). Diverging from NUHM boundary conditions, will generally gesture a deviation from SO(10) boundary conditions. If the hidden sector has representations which are not singlets under SO(10), one can expect non-trivial gaugino mass boundary conditions. So, to sum up, both NUHM and NUGM are boundary conditions which are a result of assuming SO(10) symmetric boundary conditions at the GUT scale in two disparate ways. Moreover, as can be percieved from figs presented in section IV, low energy flavor phenomenology is not much attacked by these distinct boundary conditions at high scales.
\par
In this work a SUGRA model with non universal scalar masses {\color{blue}\cite{utpal}} where the first two generations of scalar masses and the third generation of sleptons are extremely massive is also probed. This model implores the flavor changing neutral current (FCNC) issue by subscribing very large masses for the first two generations of squarks and sleptons. But the need of radiative breaking of electroweak symmetry REWSB prohibits the scalar masses from being too hugely massive. This situation is eluded by allowing third generation squark masses and the Higgs scalar mass parameters to be of small scale. {\color{blue}\cite{utpal}}. This smallness also set out to keep the naturalness problem within sovereignty.
\par
It is conventional that SUSY can be broken by soft terms of kind $- A_{0}, m_{0}, M_{1/2}$, where $A_{0}$ is the universal trilinear coupling, $m_{0}$ is the universal scalar mass, and $M_{1/2}$ is the universal gaugino mass. Stern universality amongst Higgs and matter fields of mSUGRA models can be reduced or weakened in NUHM (Non Universal Higgs Mass {\color{blue}\cite{nuhm}} models. As manifested in the results in Sec.IV in mSUGRA, the spectrum region of $M_{1/2}$ and $ m_{0} $ is found to settle in the direction of heavy side, as enabled by MEG constraints on BR($ \mu \rightarrow e \gamma $), but in NUHM, lighter spectra is feasible (owing to partial cancellations in flavor violating terms). So it enthrals one to study LFV decay $ \mu \rightarrow e \gamma $ in NUSM (Non Universal Scalar Mass Models) {\color{blue}\cite{utpal, Ga}}. As manifested in the results in sec.IV it is found that in NUSM model the gaugino masses are very massive, so as to enable very large scalar masses. In the light of the third generation squark masses and the Higgs scalar mass parameters being small, the fine tuning problem of naturalness does not get deteriorated. So to have Higgs mass around 125.9 GeV, the first two generation of squark and slepton masses as well as third generation of slepton masses habitats around 12000 GeV-16000 GeV. It is found that in NUSM model, current value of the branching fraction of the process $ \mu \rightarrow e  \gamma $: BR$(\mu \rightarrow e  \gamma )< 4.2 \times 10^{-13} $ at 90 percent C.L. allows $ tan\beta $ to lie in the region 5$-$45 and in order to have Higgs mass around 125.9 GeV, NUSM model grants $ tan\beta $ to find its appropriate value in the region 20$-$ 30 whereas in NUHM model it is seen that $ tan\beta $ occupies itself around 7$-$13 for Higgs mass around 125.9 GeV. 
\par
Therefore it is conceived that indication of LFV could be tested at high luminosity HE/HL run of LHC, if SUSY sparticles are spotted around TeV range. It is eminent that, no SUSY partner of SM has been noticed yet at LHC, and this tipst to a high scale SUSY theory. The LHC has rigorous limits on sparticles, which could entail a tuning of EW symmetry at a few percent level {\color{blue}\cite{T.Gherg}}-{\color{blue}\cite{J.Fan}}. Thus some substitute to low scale SUSY theories have been recommended. Few of them are $ - $ minisplit SUSY {\color{blue}\cite{Villadoro}} and maximally natural SUSY {\color{blue}\cite{Savas}}. In minisplit SUSY the scalar sparticles are massive than the sfermions (gauginos and higgsinos), so that sfermions could be spotted at HE-LHC. Scalar sparticles could be available everywhere in the range (10$-10^{5})$ TeV. In maximally natural SUSY, the 4D theories rises from 5D SUSY theory, with  Scherk-Schwarz SUSY breaking at a Kaluza-Klein scale $ \sim \frac{1}{R}$ of several TeV {\color{blue}\cite{Savas}}. Some prospects of LFV in those theories have been explored in {\color{blue}\cite{Isabel}}. 

The paper has been organised as follows. In section II, connections of LFV with type II See Saw mechanism in $ S_{4} \times Z_{n} $ flavor symmetric SUSY SO(10) theory is discussed. In section III, the values of various parameters used in this analysis has been presented. The software SuSeFLAV {\color{blue}\cite{Garani}} is used to determine BR($ \mu \rightarrow e \gamma $). Section IV manifests itself in results and their analysis. Section V outlines the work. 

\section{Lepton Flavor Violation $ \mu \rightarrow e \gamma $ decay in $ S_{4} \times Z_{n} $ flavor symmetric SUSY SO(10) theory.}
\label{sec:1}
\subsection{\textbf{Lepton Flavor Violation and observables.}}
\label{sec:2}
In SUSY theory, non-diagonal mass matrix elements in the slepton mass matrix are the initiator of Lepton Flavor Violation processes. 
\begin{center}
\begin{figure*}[htbp]
\includegraphics[height=5.9cm,width=10cm]{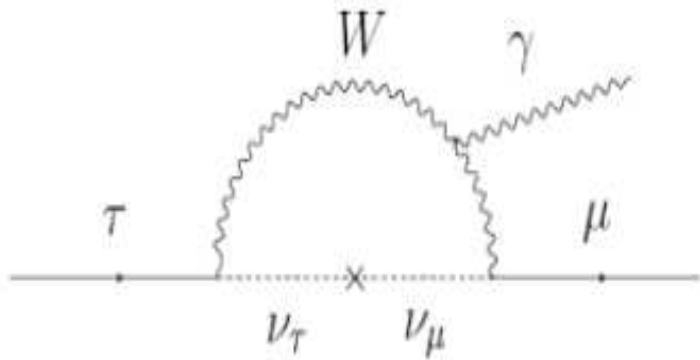}

\caption{Examples of Feynmann Diagrams contributing to $ \tau \rightarrow \mu +\gamma $ processes in SUSY models. }
\label{fig:1}
\end{figure*}
\end{center}

\begin{center}
\begin{figure*}[htbp]
\includegraphics[height=6cm,width=15cm]{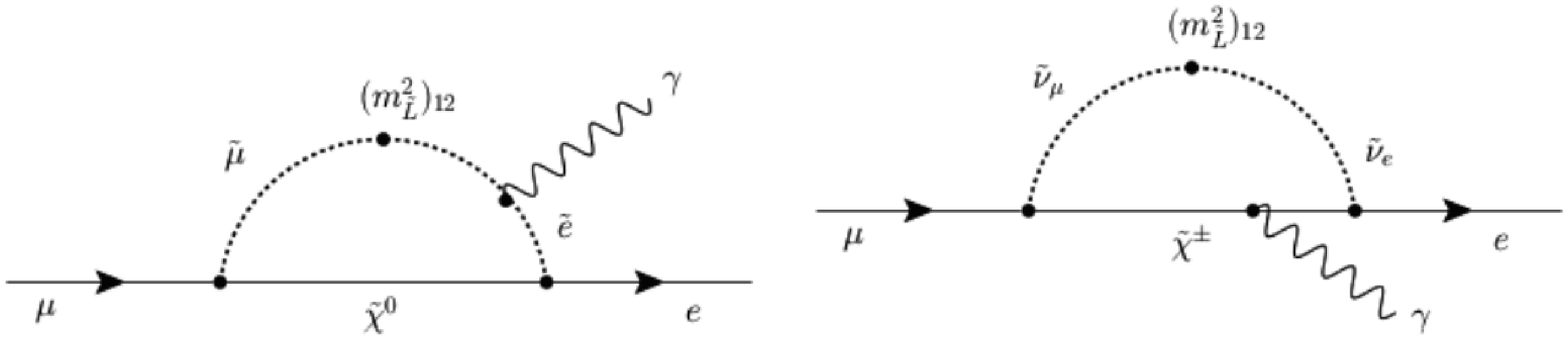}

\caption{Examples of Feynmann Diagrams contributing to $ \mu \rightarrow e +\gamma $ processes in SUSY models. }
\label{fig:1}
\end{figure*}
\end{center}
The SUSY SO(10) theory obviously embraces the see–saw mechanism. The existence of heavy RH neutrinos at an intervening scale give on to the running and originate flavor violating entries in the left-handed slepton mass matrix at the weak scale {\color{blue}\cite{L.Cabbibi2012}}. The lepton flavour violating entries in the SO(10) SUSY GUT framework can be percieved in terms of the low energy parameters. In the mass insertion (MI) method with leading log approximation,
the branching fractions for different LFV processes $ l_{i} \rightarrow l_{j} + \gamma $ can be resembled as
\begin{equation}
\text{BR} \left( l_{i} \rightarrow l_{j}+\gamma \right)\approx \alpha^{3}\frac{\vert (m^{2}_{\bar(L)})(LL)_{ij}\vert^{2}}{G_{F}^{2}m^{8}_{SUSY}}\tan^{2}\hspace{.01cm}\beta \linebreak \text{BR} \left( l_{i}  \rightarrow  l_{j}\nu_{i}\tilde{\nu_{j}} \right)
\end{equation}
where $ M_{SUSY } $ is mas scale of the SUSY particles, $ \alpha $ is the fine structure constant and $ G_{F} $ is the Fermi constant. The 6 $\times$ 6 slepton mass matrix is described as 
\begin{equation}
m^{2}_{\bar(L)} = \begin{pmatrix}
m^{2}_{\bar(L)})(LL)_{ij}& m^{2}_{\bar(L)})(LR)_{ij}\\
m^{2}_{\bar(L)})(RL)_{ij} & m^{2}_{\bar(L)})(RR)_{ij}\\
\end{pmatrix}
\end{equation}
where \textit{LL, RL, LR, RR} are 3$ \times $ 3 entries established on the chirality tag of sfermions.
\par 
The leading log entries approximation in mSUGRA is defined as {\color{blue}\cite{A.Faccia}}
\begin{equation}
 \left( m^{2}_{\tilde{L}} \right)_{i\neq j} = \frac{-3m_{o}^{2}+A_{o}^{2}}{ 8\pi^{2}} \sum_{k}
      \left(f_{\nu}^{\star}\right)_{ik}\left(f_{\nu}\right)_{jk} \\ log\left(\frac{M_{X}}{M_{R_{k}}}\right) 
\end{equation}
 where $M_{X}$ is the GUT scale, $M_{R_{k}}$ is the $k^{th}$ heavy RH majorana neutrino scale,  $m_{0}$ and $A_{0}$ are universal soft mass and trilinear terms at the high GUT scale. $ f_{\nu} $ are the Dirac neutrino Yukawa (DNY) couplings. The flavour violation is specified in terms of the quantity $\delta_{ij}=\frac{\Delta_{ij}}{\overline{m}^{2}_{\tilde{l}}}$, where ${{\overline{m}}}^{2}_{\tilde{l}} $ is the geometric mean of the slepton squared masses {\color{blue}\cite{F.Gabbiani}}, and $ \Delta_{i \neq j}$ are flavour non diagonal entries of the slepton mass matrix instigated at the weak scale due to RG evolution. The mass insertions are labelled into the LL/LR/RL/RR kinds {\color{blue}\cite{Masina}}, based on the chirality of the SM fermions. The fermion masses are formed by renormalisable Yukawa couplings of the 10$\oplus$126$\oplus\overline{120}$ portrayal of scalars of SO(10) GUTs. The Dirac neutrino Yukawa couplings $ f_{\nu} $ at the high GUT scale in $ S_{4} \times Z_{n} $ flavor symmetric SUSY SO(10) theory is used in this work from {\color{blue}\cite{bd}} .
 \begin{equation}
 f_{\nu}=\frac{1}{\upsilon sin\beta}M_{D}
 \end{equation}
 Here $M_{D}$ is the Dirac neutrino mass matrix. The off-diagonal flavor violating entries at the weak scale in eq. {\color{blue}(3)} are then absolutely decided by using $f_{\nu}$ from eq. {\color{blue}(4)}. The $ \delta $s from the RGEs are calculated, using the leading log approximation. The soft SUSY masses are flavour universal at the input scale, so, off diagonal entries in the LL sector are developed by running of right handed neutrinos in the renormalisation group equation loops. For using the leading log expression (eq. {\color{blue}(3)}) one requires the mass of the heaviest right handed neutrino, which is here, $\sim 10^{16} $ GeV. The induced off-diagonal leading log entries pertinent to LFV decay $l_{i}$ $\rightarrow$ $l_{j}$ +$ \gamma $ are of the magnitude of (setting down $ A_{0} $ to 0){\color{blue}\cite{Ga}}
\begin{equation}
\left( \delta_{LL}\right) _{\mu e} = \frac{-3}{8\pi^{2}} \left( f_{\nu}^{\star}\right) _{13}\left( f_{\nu}\right) _{23}ln\left( \frac{M_{X}}{M_{R_{3}}}\right) 
\end{equation}

\begin{equation}
\left( \delta_{LL}\right) _{\tau \mu} = \frac{-3}{8\pi^{2}} \left( f_{\nu}^{\star}\right) _{23}\left( f_{\nu}\right) _{33}ln\left( \frac{M_{X}}{M_{R_{3}}}\right) 
\end{equation}

\begin{equation}
\left( \delta_{LL}\right) _{\tau e} = \frac{-3}{8\pi^{2}} \left( f_{\nu}^{\star}\right) _{13}\left( f_{\nu}\right) _{33}ln\left( \frac{M_{X}}{M_{R_{3}}}\right) 
\end{equation}

\begin{table}
\caption{Governing values of $\delta_{ij}$ that enter eq. {\color{blue}(5,6,7)}
for in $ S_{4} \times Z_{n} $ flavor symmetric SUSY SO(10) theory.}
\label{tab:2}       % Give a unique label
\begin{tabular}{ll}
\hline\noalign{\smallskip}
\textbf{LFV contributions} &\textbf{ For $ S_{4} \times Z_{n} $ flavor symmetric SUSY SO(10) theory case} \\
\noalign{\smallskip}\hline\noalign{\smallskip}
$\delta_{12}$ & $0.6672\times 10^{-4}$ \\ 
 
$\delta_{23}$ & $1.5634\times 10^{-4}$ \\ 
 
$\delta_{31}$ & $0.7377\times 10^{43}$ \\ 

\noalign{\smallskip}\hline
\end{tabular}
\end{table}
\par 
 In NUHM models, the expression
 $(-3m_{o}^2+A_{o}^2)$ of mSUGRA models in {\color{blue}eq.(3)} is put back by the leading log approximation for the slepton mass matrix element that induces the process $ \mu\rightarrow e +\gamma $, as $(-2m_{o}^2+A_{o}^2+m^2_{H_u})$. $ m_{H_u} $ is the  soft mass terms of the up type Higgs at the high scale. At the GUT scale, the NUHM case, 
$
m_{H_{u}}=m_{H_{d}}
$
Besides, due to a relative sign difference between the universal soft mass terms for the matter fields and the Higgs mass terms at the GUT scale, cancellations occurs for
$m^{2}_{H_{u}}\approx -2m_{0}^{2}$, or increment for $m^{2}_{H_{u}} \geq m_{0}^{2}$
in contrast to mSUGRA for the off diagonal flavor violating entries at the weak scale.

\subsection{\textbf{$ S_{4} \times Z_{n} $ flavor symmetry}}
\label{sec:2}

Dirac Neutrino Yukawa couplings from a supersymmetric SO(10) grand unified theory (GUT) of flavor relevant to an $ S_{4} $ family symmetry is used here in this work. It makes use of the fact that, SO(10) theory combined with type II seesaw mechanism for generating  neutrino masses mingled with a simple assertion that the dominant Yukawa coupling matrix (the 10-Higgs coupling to matter) has rank one. The rank one model arises within some reasonable speculation as a constructive field theory from vectorlike 16 dimensional matter fields with masses lying above the GUT scale. $ S_{4} $ flavon multiplets
get vevs in the ground state of the theory. By enlarging the $ S_{4} $ theory with
an additional discrete symmetry $ Z_{n}  $, it has been found that the flavon vacuum field alignments acquire a discrete values of parameters assuming that some of the higher dimensional couplings are small. An observed set of vacuum alignments directs one to an unification of quark-lepton flavor: (i) the lepton mixing matrix that is dominantly tri-bimaximal with small corrections related to quark mixings; (ii) quark lepton mass is related at GUT scale as $ m_{b} = m_{\tau} $and $m_{\mu} = 3m_{s}$ and (iii) the solar to atmospheric neutrino mass ratio $m_{solar}/m_{atm} = \theta $ Cabibbo, which in tally with the experiments.

The mass matrix becomes 
\begin{equation}
M_{\nu} = \begin{pmatrix}
0 & c & c\\
c & a & c-a\\
c & c-a & a\\
\end{pmatrix}
\end{equation}
where $\frac{c}{a} = \lambda \leq 1$. It is diagonalized by the tri-bi-maximal matrix,
\begin{equation}
U_{TB} = \begin{pmatrix}
\sqrt\frac{2}{3} & \sqrt\frac{1}{3}& 0\\
-\sqrt\frac{1}{6} & \sqrt\frac{1}{3} & - \sqrt\frac{1}{2}\\
-\sqrt\frac{1}{6} & \sqrt\frac{1}{3} & \sqrt\frac{1}{2}\\
\end{pmatrix}
\end{equation} 
This is  not the full $U_{PMNS}$ matrix which is going to get small radiative corrections from diagonalization of the charged lepton mass matrix, which generates small reactor angle, $ \theta_{13} $ along with small $ \theta_{\odot} $ and $ \theta_{atm} $.
The neutrino masses are given by $ m_{\nu_{3}} \equiv 2a -c$ ; $m_{\nu_{2}} = 2c$ and $m _{\nu_{1}} = -c$.
To fit observations, $ \frac{c}{a} \simeq \Delta m^{2}_{\odot}/\Delta m^{2}_{atm} \sim 0.2$, which realises the neutrino masses to be, $m_{\nu_{3}} = 0.05 eV$, $m_{\nu_{3}} = 0.01 eV$, and $m_{\nu_{1}} $ = 0.005 eV. 
\section{Branching Ratio, BR($\mu \rightarrow e\gamma$) in CMSSM, NUHM, NUSM}
In this section the computations on the  LFV constraints of BR($\mu \rightarrow e\gamma$) in $ S_{4} \times Z_{n} $ flavor symmetric SUSY SO(10) theory  with type II Seesaw mechanism using the NUHM, CMSSM, NUSM like boundary conditions through detailed numerical analysis is presented. The soft parameter space for CMSSM in the following ranges is studied.
\begin{equation*}
tan\beta \in \left[ 1, 60\right]
\end{equation*}
\begin{equation*}
 \Delta m_{H} \in 0 
\end{equation*}
\begin{equation*}
 m_{0} \in \left[ 0, 8\right] \hspace{.1cm}\text{TeV} 
\end{equation*}
\begin{equation*}
 M_{1/2} \in \left[ 0.3, 4.5\right] \hspace{.1cm} \text{TeV}
\end{equation*}
\begin{equation*}
 A_{0} \in \left[ -3m_{0} , +3m_{0} \right]
\end{equation*}
\begin{equation}
sgn\left( \mu\right) \in\lbrace-,+\rbrace 
\end{equation}
\vspace{.05cm}
The analytical part is done using the publicly available package SuSeFLAV {\color{blue}\cite{Garani}}. LFV for the non universal Higgs model without completely universal soft masses at high scale is scrutinised. Range of examination of various SUSY parameters, used in NUHM are:

\begin{equation*}
30\hspace{.1cm} \text{GeV} \leq m_{0} \leq 8\hspace{.1cm} \text{TeV} 
\end{equation*}
\begin{equation*}
30\hspace{.1cm} \text{GeV} \leq M_{1/2} \leq 5\hspace{.1cm}\text{TeV}
\end{equation*}
\begin{equation*}
- 9.5 \hspace{.1cm}\text{TeV} \leq m_{H_{u}} \leq +9.5 \hspace{.1cm}\text{TeV}
\end{equation*}
\begin{equation*}
-9.5\hspace{.1cm}\text{TeV} \leq m_{H_{d}} \leq +9.5 \hspace{.1cm}\text{TeV} 
\end{equation*}
\begin{equation}
-24 \hspace{.1cm}\text{TeV} \leq A_{0} \leq + 24\hspace{.1cm}\text{TeV}
 \end{equation}
The $\Delta^{LL}_{i\neq j}$  owing to non universal Higgs  and $ m_{h} $ $\geq$ 125 GeV puts a powerful restraint on SUSY parameter space. Due to partial cancellations in the entrance of $\Delta^{LL}_{i\neq j}$ in NUHM case, a substantial region of parameter space can be probed by MEG.
\par 
Contingent scans for the following range of parameters in NUSM model is performed {\color{blue}\cite{utpal}} and the SUSY particle spectrum using the publicly available package SuSeFLAV {\color{blue}\cite{Garani}} is created. 
\begin{equation*}
tan\beta \in \left[ 5, 60\right] 
\end{equation*}

\begin{equation*}
 m_{0} \in \left[ 0, 16\right] \hspace{.1cm}\text{TeV} 
\end{equation*}
\begin{equation*}
 M_{1/2} \in \left[ 0, 6\right] \hspace{.1cm}\text{TeV} 
\end{equation*}
\begin{equation*}
 A_{0}  \in 0\hspace{.1cm} \text{TeV} 
\end{equation*}
\begin{equation}
m_{H_{u}}=m_{H_{d}} \in  0\hspace{.1cm} \text{TeV} 
\end{equation}
Massive right handed neutrinos used in our calculations are - $ M_{R_{1}}=10^{13} \hspace{.1cm}\text{GeV}$, $M_{R_{2}} = 10^{14}\hspace{.1cm}\text{GeV}$, and $M_{R_{3}}=  10^{16}\hspace{.1cm}\text{GeV}$.
The central values of $ \Delta m^{2}_{sol} $, $\Delta m^{2}_{atm}$ and $\theta_{13}$, from the recent global fit of neutrino data {\color{blue}\cite{tor}} is employed. In Table 1 the presiding values of $\delta_{ij}$ that enter {\color{blue}eq.(5,6,7)} is presented.
\section {Calculations and Discussion on Results}

In this section, study on the computation of results presented in section 3 is discussed.
\subsection{\textbf{Complete Universality - CMSSM }}
\label{sec:2}
At the high scale, the parameters of the CMSSM model described are $m_{0}$, $A_{0}$ and unified gaugino mass $ M_{1/2} $. Also there are the Higgs potential parameter $ \mu $ and the ratio of the Higgs VEVs, tan$\beta$. The overall SUSY mass spectrum is estimated once those parameters are accessible. The updated MEG constraint {\color{blue}\cite{17,44}} set together with a big $\theta_{13}$ {\color{blue}\cite{tor}} brings forth significant restrictions on SUSY parameter space in CMSSM. As depicted from fig {\color{blue}3a}, few part of the paramater space is allowed for tan $ \beta $ = 5$-$ 60 in CMSSM as constrained by future MEG limit for BR($ \mu \rightarrow e \gamma $) which is 6$ \times 10^{-14}$. So to conclude it is seen that  the parameter space $M_{1/2} \geq $ 10 GeV is permitted by present MEG bounds on BR($ \mu \rightarrow e \gamma $), incidentally future MEG limit prohibits strictly almost whole $M_{1/2}$ space. Similarly as seen from fig {\color{blue}3b}, the current MEG limit allows very heavy spectra for $ m_{0} $, which favours $ m_{0} $ to lie between 4.5 TeV to 8 TeV which is indeed massive. The permitted space in  fig {\color{blue}4a} needs very massive spectra, i.e. $ m_{0} $ $ \geq $ 6 TeV and it mostly lies around 8 TeV for the whole spectrum of $M_{1/2}$. From fig {\color{blue}4b}, it is found that the current bound for the process BR($ \mu \rightarrow e \gamma $) $4.2 \times 10^{-13}$ sets very stringent constraint on tan $ \beta $, which find its value from around 2 to 10. $ tan\beta \leq $ 2 and $ tan\beta \geq $ 10 is ruled out in the CMSSM case. In  fig {\color{blue}4c, 4d} the light Higgs mass, $ m_{h} $ as a function of $ m_{0} $, $M_{1/2}$ in the CMSSM case is computed. The range of Higgs mass as given by the data at LHC, i.e 123 $\text{GeV} \le m_{h} \le$ 127 $\text{GeV}$ allows $m_{0} $ 6 TeV $\geq$ 8 TeV as restricted by constrained stringent MEG bounds on BR($ \mu \rightarrow e \gamma $). The region $M_{1/2} \geq$ 3 TeV is permitted as can be seen from fig {\color{blue}4d}. The asymmetry in $ A_{0} $ GeV can be seen from fig {\color{blue}4g}. To have Higgs mass around, 123 $\text{GeV} \le m_{h} \le$ 127 $\text{GeV}$, the values of $ tan\beta $ from 3 to 6 are mostly allowed. The dependancy of $ A_{0} $ on $ tan \beta $ is shown in fig {\color{blue}4g} 

\begin{center}
\begin{figure*}[htbp]
\centering{
\begin{subfigure}[]{\includegraphics[height=5.9cm,width=7.9cm]{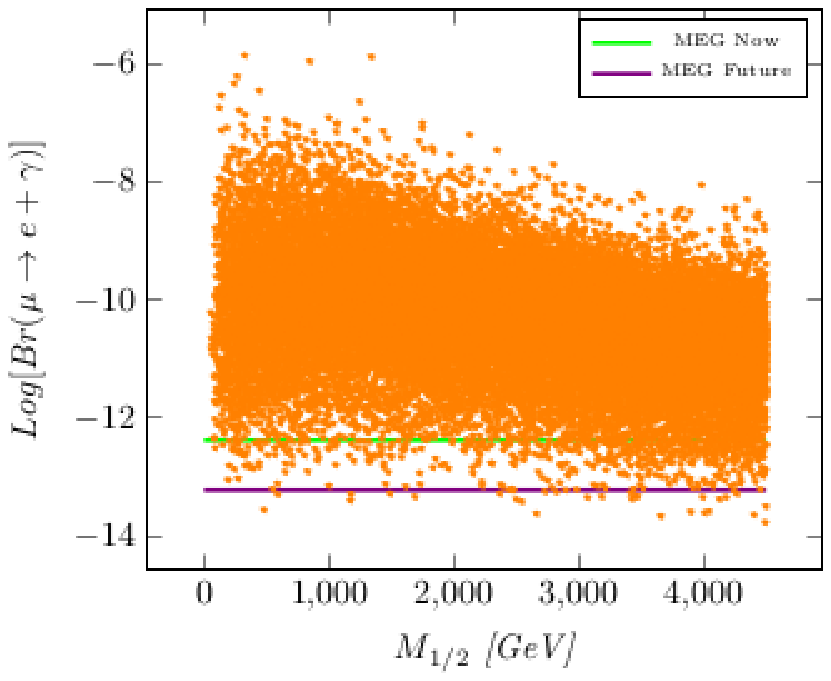}}\end{subfigure}
\begin{subfigure}[]{\includegraphics[height=	5.93cm,width=8cm]{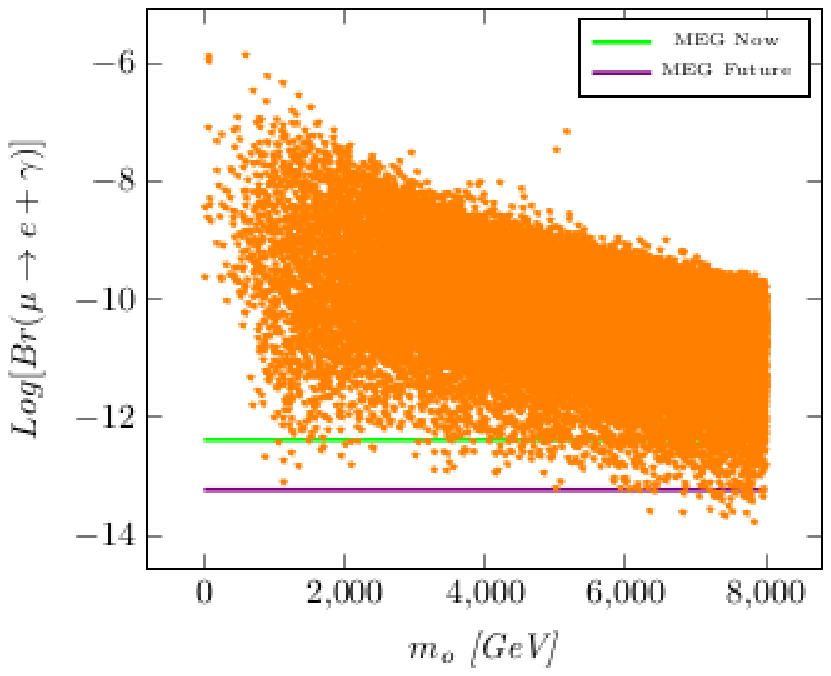}}\end{subfigure}\\

\caption{The outcome of the calculations are presented for CMSSM case. In fig {\color{blue}3a, 3b}, different
 horizontal lines depicts the present (MEG 2016) and future MEG constraints for BR($ \mu $ $ \rightarrow $ e + $ \gamma $).}}
\label{fig:1}
\end{figure*}
\end{center}

\begin{center}
\begin{figure*}[htbp]
\centering{
\begin{subfigure}[]{\includegraphics[height=4.9cm,width=7.9cm]{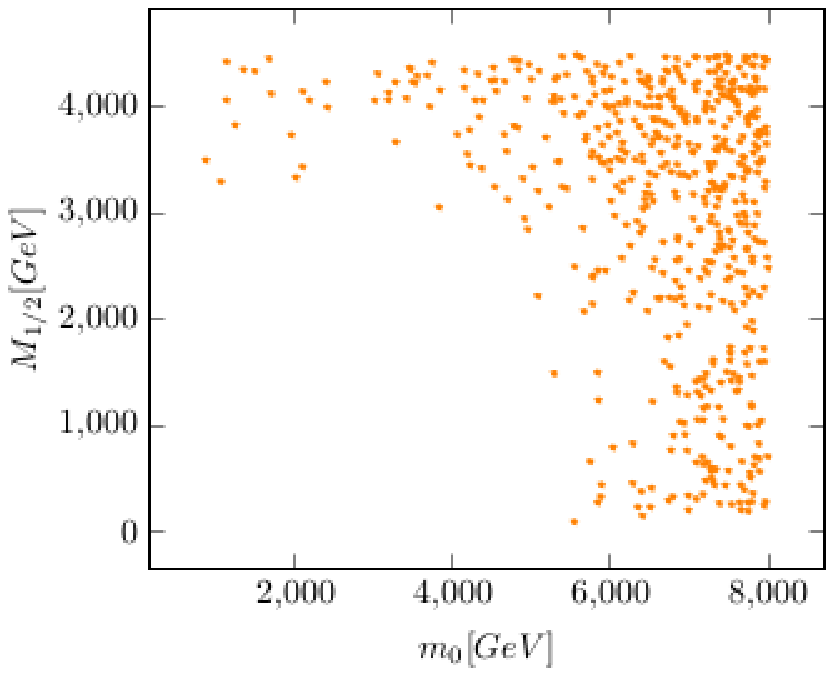}}\end{subfigure}
\begin{subfigure}[]{\includegraphics[height=4.9cm,width=8cm]{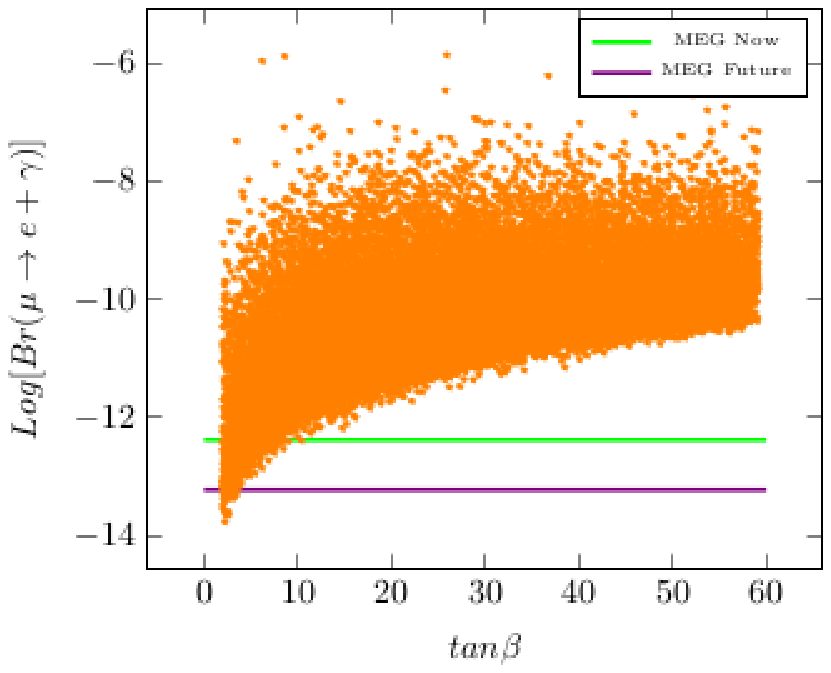}}\end{subfigure}\\
\begin{subfigure}[]{\includegraphics[height=5.9cm,width=7.7cm]{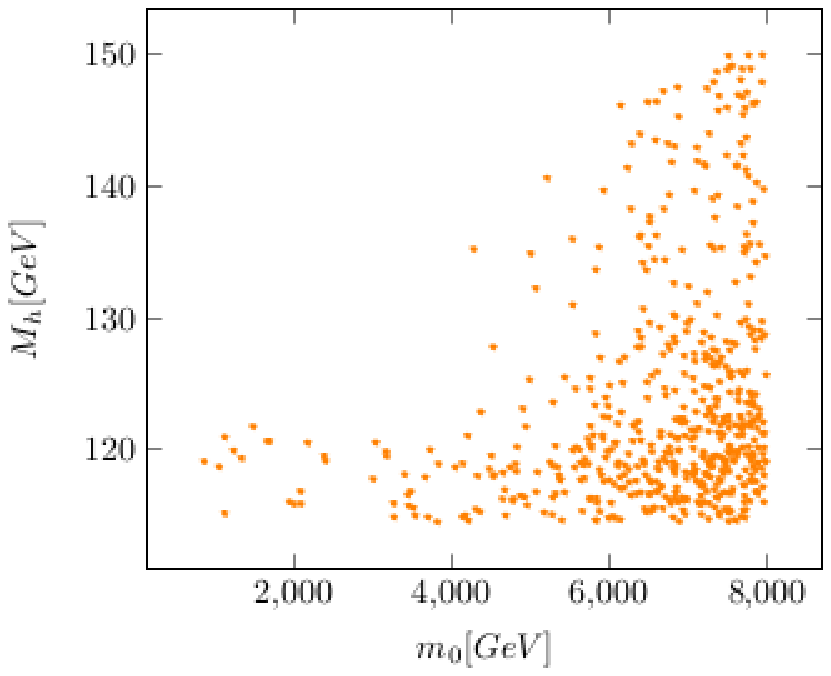}}\end{subfigure}
\begin{subfigure}[]{\includegraphics[height=5.9cm,width=7.9cm]{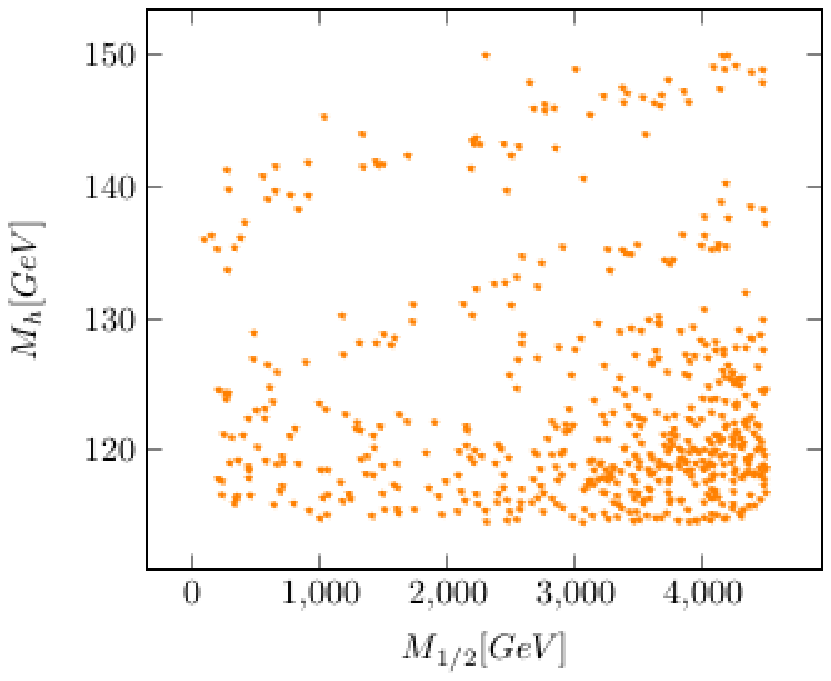}}\end{subfigure}\\
\begin{subfigure}[]{\includegraphics[height=5.9cm,width=7.7cm]{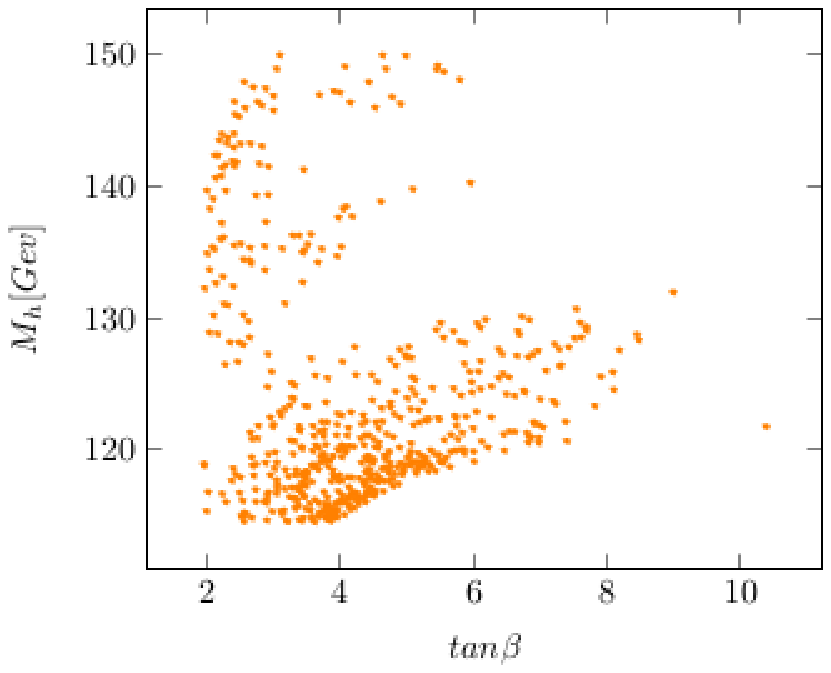}}\end{subfigure}
\begin{subfigure}[]{\includegraphics[height=5.9cm,width=7.9cm]{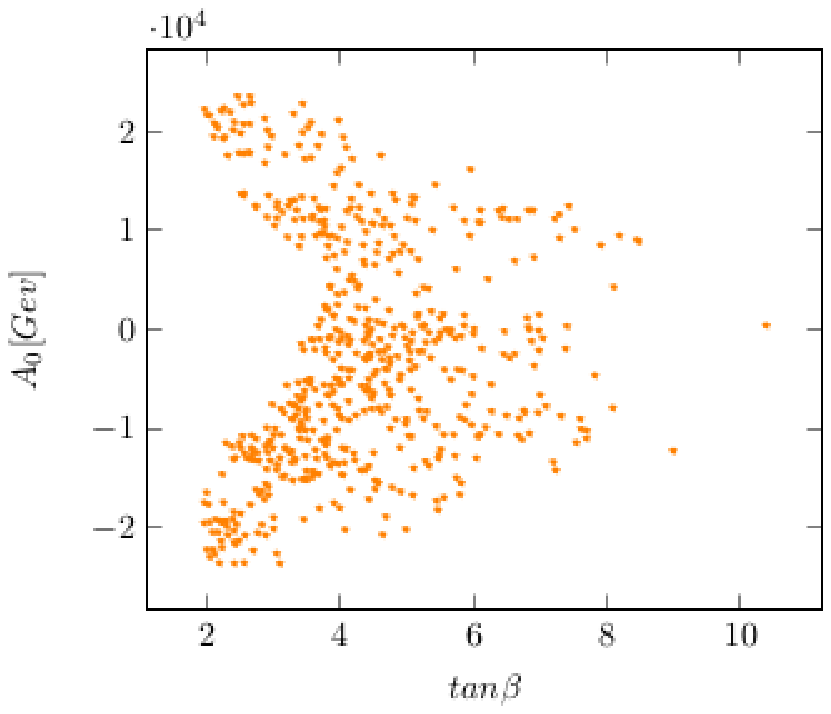}}\end{subfigure}\\
\begin{subfigure}[]{\includegraphics[height=4.9cm,width=7.9cm]{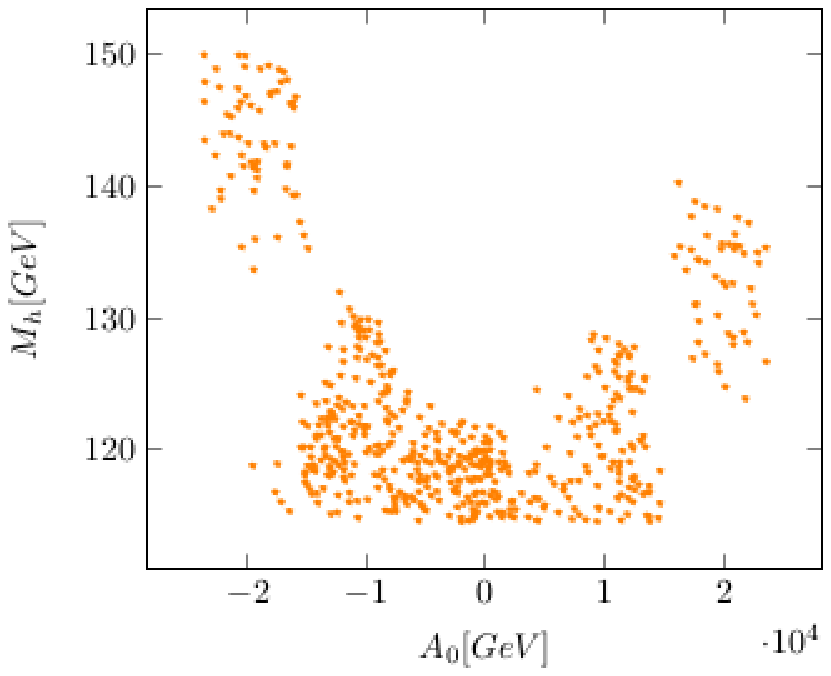}}\end{subfigure}\\
\caption{ In figs {\color{blue}(4a-4e)} allowed SUSY parameters region as constrained by MEG 2016 bound is presented.}}
\label{fig:1}
\end{figure*}
\end{center}

\subsection{\textbf{Non Universal Higgs Model (NUHM)}}
The cMSSM+RHN model parameter set is investigated in the literature {\color{blue}\cite{l,m,n}}. Also, the NUHM1 +RHN models were extensively studied for specific non-universal scenarios with the following GUT-scale mass relations: NUHM1 ($m^{2}_{H_{u}} = m^{2}_{H_{d}} = m^{2}_{20} $) in {\color{blue}\cite{nuhm}}
The leading log approximation for the slepton mass matrix element that induces the process $ \mu\rightarrow e +\gamma $ is 
 
 \begin{equation}
 \left( m^{2}_{\tilde{L}} \right)_{i\neq j} = \frac{-2m_{o}^{2}+A_{o}^{2}+m^{2}_{H_{u}}}{ 8\pi^{2}} \sum_{k}
      \left(f_{\nu}^{\star}\right)_{ik}\left(f_{\nu}\right)_{jk} \\ log\left(\frac{M_{X}}{M_{R_{k}}}\right) 
\end{equation}
For low values of $ M_{1/2} $, the approximation holds good but for values of $ M_{1/2} \simeq $ 1 TeV, the branching
fractions may be up to a factor of 10 dissimilar than the evaluation from RGE running. If the hierarchy of dirac neutrino yukawa couplings are alike to that of the up-type in the Standard Model, where the third generation predominantly
dominates, then the largest donation is from the $k = 3$ terms in the summation. Calculation of Log[BR($ \mu \rightarrow e+\gamma$)], in the NUHM case leads to a good approximation for a light/pre-LHC SUSY mass spectrum due to the cancellations between the soft SUSY parameters in {\color{blue}eq.(13)}. Alongside, the consequences of the results accessed in NUHM case is discussed. In fig. {\color{blue}5a} $m_{0}$ [GeV] Vs Log[BR($ \mu \rightarrow e+\gamma$)], and the fig. {\color{blue}5b} in the right panel depicts $M_{1/2}$ [GeV] Vs Log[BR($ \mu \rightarrow e+\gamma$)]. Various horizontal lines in fig. {\color{blue}5a, 5b} correspond to present and future bounds on BR($ \mu $ $ \rightarrow $ e + $ \gamma $). It is see from the figs. {\color{blue}5a} and {\color{blue}5b} that due to the partial cancellations between the NUHM parameters $ m_{0} $, $ A_{0} $ and $ m_{H_{u}} $, almost all of the NUHM parameter space is going to be probed by the stringent MEG bounds for a good approximation of a light/pre-LHC SUSY mass spectrum.

\begin{center}
\begin{figure*}[htbp]
\centering{
\begin{subfigure}[]{\includegraphics[height=5.9cm,width=7.9cm]{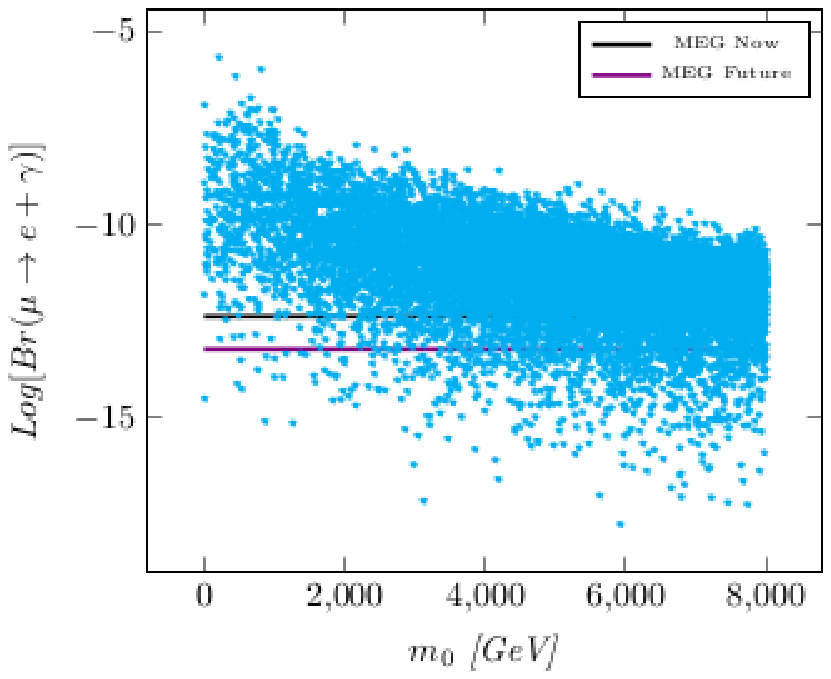}}\end{subfigure}
\begin{subfigure}[]{\includegraphics[height=6cm,width=8cm]{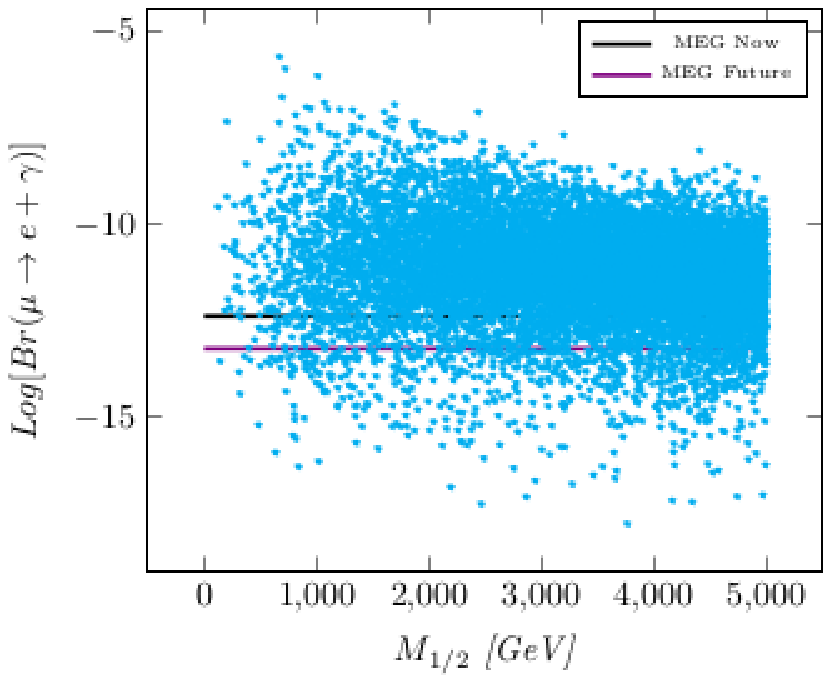}}\end{subfigure}\\
\begin{subfigure}[]{\includegraphics[height=5.9cm,width=7.9cm]{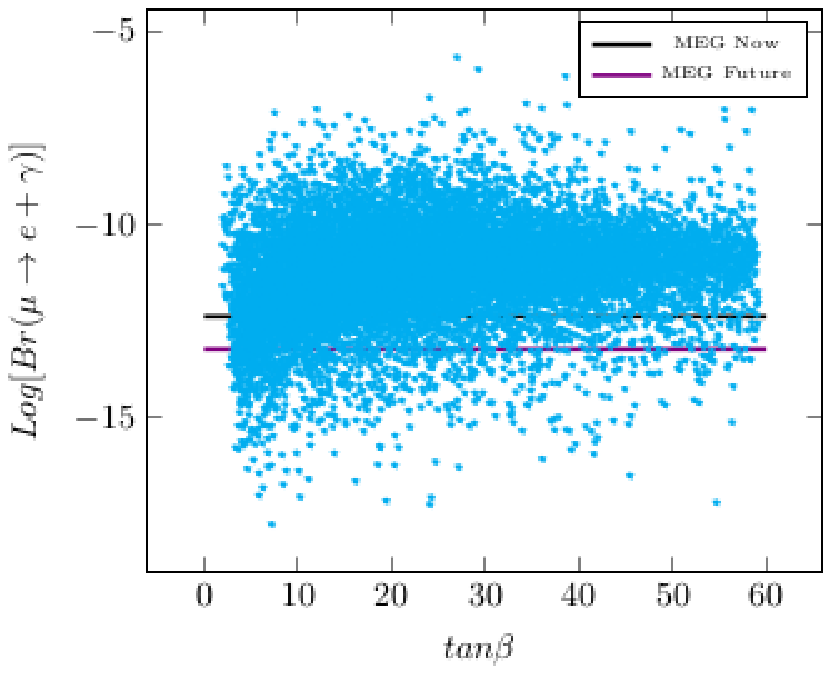}}\end{subfigure}
\begin{subfigure}[]{\includegraphics[height=6cm,width=8cm]{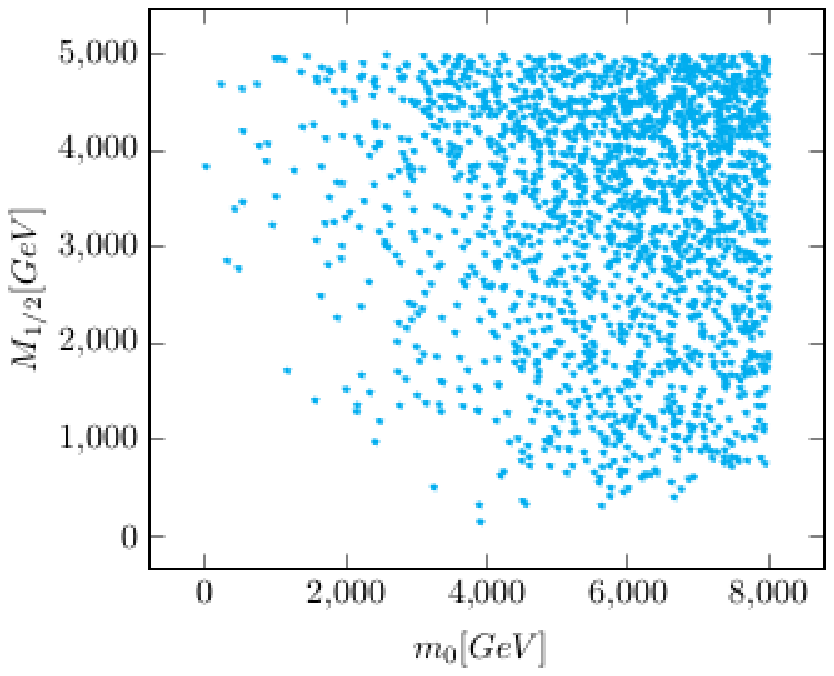}}\end{subfigure}\\

\caption{The consequences of the analysis and calculations are presented for NUHM case. In fig {\color{blue}5a,5b}, different
 horizontal lines illustrates the present (MEG 2016) by MEG Collaboration and future MEG bounds for BR($ \mu $ $ \rightarrow $ e
 + $ \gamma $). Figs. {\color{blue}5c,5d} portrays the allowed SUSY parameter space for different parameters, as is constrained by stringent by MEG 2016 bounds.}}
\label{fig:2}
\end{figure*}
\end{center}
In figs. {\color{blue}5c}, Log[BR($ \mu \rightarrow e+\gamma$)] Vs $ tan\beta $ is presented. It is grasped from the fig. {\color{blue}5c}, that for present bound of MEG, almost all values of $ tan\beta $ from 5 to 60 are favoured, whereas future MEG bound restricts the values of $tan\beta$ to 5 $ - $ 40. Fig.{\color{blue}5d} represents $ m_{0} $ Vs $ M_{1/2} $. The SUSY parameter space $M_{1/2} - m_{h} $ and $m_{0} - m_{h} $ is presented, as allowed by present MEG bounds in figs. {\color{blue}6a,6c}. For Higgs mass  to be around 126 GeV, values of $ M_{1/2} $ from 4 TeV to 5 TeV are mostly allowed. Similarly for $ m_{h} $ around 126 GeV, region 6 TeV $ \leq m_{0}\leq $ 8 TeV are mostly allowed. In $ \delta^{LL}_{i \neq j} $ owing to the  cancellations between $ m^{2}_{H_{u}} $ and $ m_{0}^{2} $ a huge susy soft parameter space compared to CMSSM is permitted which would be easily spotted at the HE LHC fulfilling the present cLFV constraints measured by MEG collaboration in 2016. Fig {\color{blue}6b} shows $A_{0}$ [GeV] Vs $ m_{h} $ [GeV]. Negative values of $A_{0}$ are favoured in order to have Higgs mass around 126 GeV. Fig {\color{blue}6d} represents tan $ \beta $ Vs $ m_{h} $. The last row in the right panel depicts the constraintor restriction on tan$ \beta $. Amost all values of  tan$ \beta$ from around 5 to 40 are allowed. In the CMSSM case, the resrictions in the low tan$ \beta $ are due to $ m_{h} $ while  those at high tan β are from the constraints on BR($ \mu \rightarrow e+\gamma$). In the CMSSM case, a light
Higgs mass around 125 GeV does not necessarily implies a suppressed flavor violating entry instead of the largeness of A-terms required. In fact flavor violation constraints
are indeed still very powerful.
\begin{center}
\begin{figure*}[htbp]
\begin{subfigure}[]{\includegraphics[height=5.9cm,width=7.9cm]{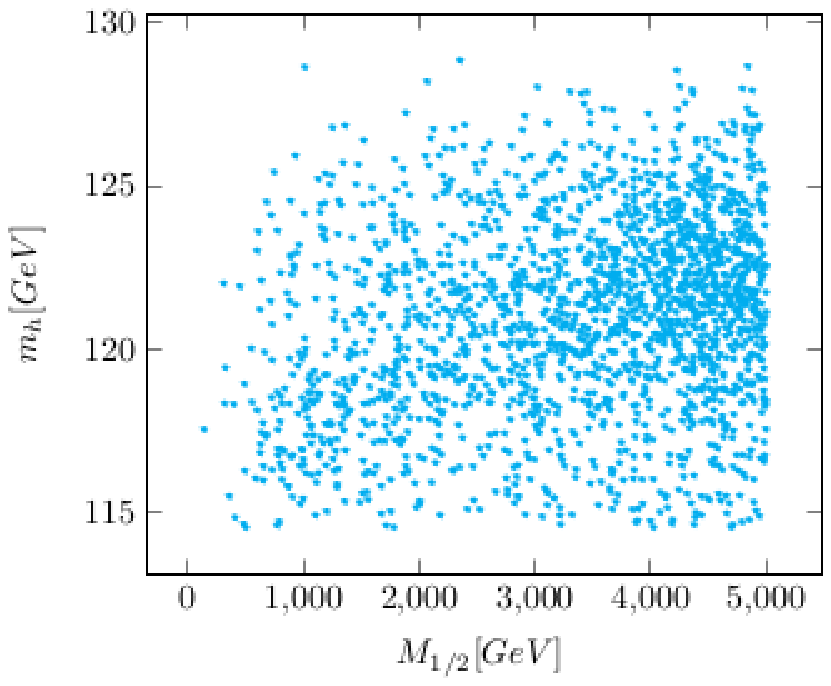}}\end{subfigure}
\begin{subfigure}[]{\includegraphics[height=6cm,width=8cm]{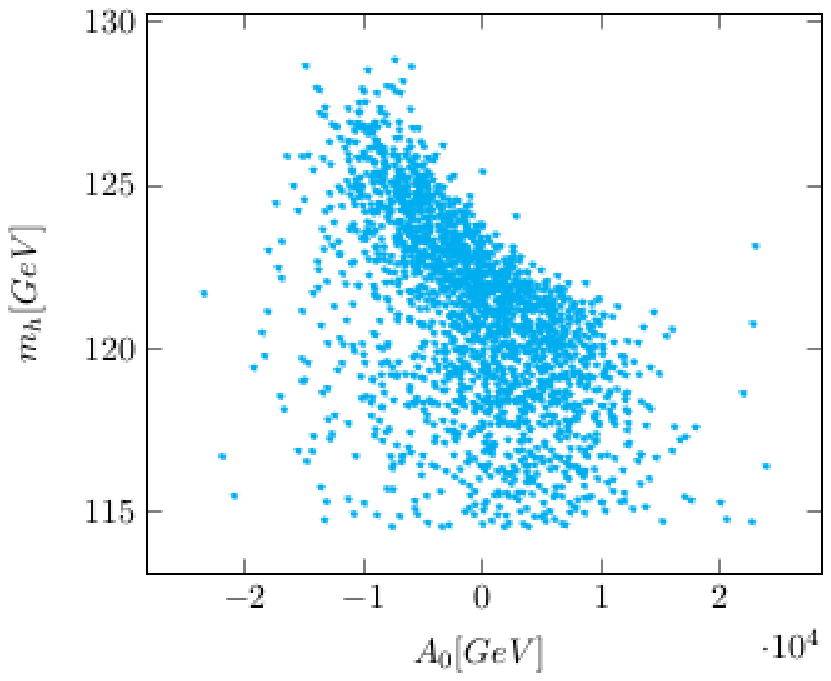}}\end{subfigure}\\
\begin{subfigure}[]{\includegraphics[height=5.9cm,width=7.9cm]{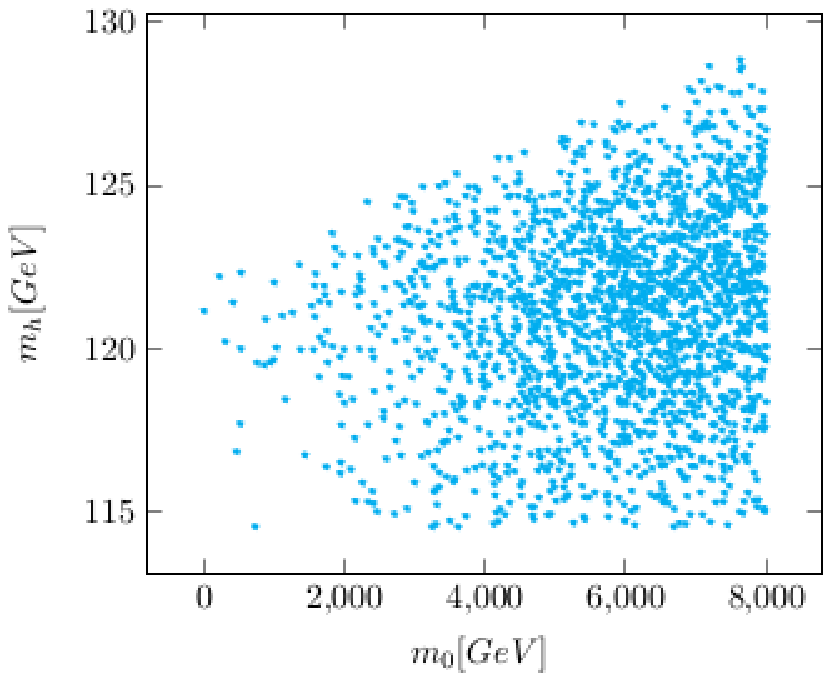}}\end{subfigure}
\begin{subfigure}[]{\includegraphics[height=5.9cm,width=7.9cm]{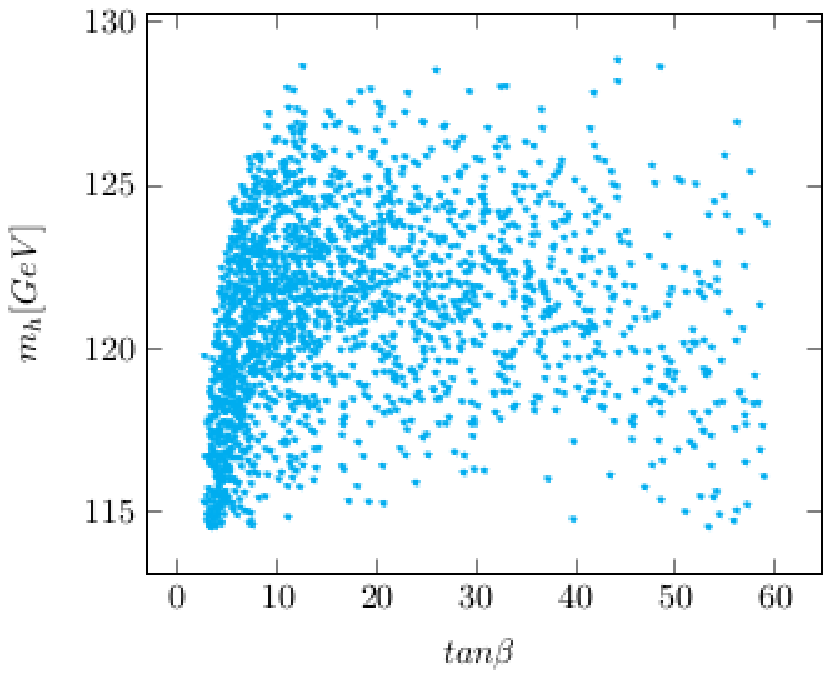}}\end{subfigure}\\
\caption{Figs. {\color{blue}6a-6d} describe the allowed SUSY region for different soft SUSY parameter space, as is obstructed by stringent MEG 2016 bounds.}
\label{fig:2}
\end{figure*}
\end{center}

\subsection{\textbf{Non Universal Scalar Mass Models (NUSM)}}

The parameters of NUSM model is given by {\color{blue}\cite{Ga}}
\par 
$\hspace{.7cm}$ $\text{tan}\beta$, $M_{1/2}$, $A_{0}$, $\text{sgn}(\mu)$, \text{and} $m_{0}$. 
\newline
The parameters of this model have the exact role to those in CMSSM case except for a major dissimilarity in the scalar sector. First two generations scalars masses (squarks and sleptons) and the third generations sleptons masses are designated as $ m_{0} $ at the GUT scale. Here $ m_{0} $ is  streches over a very large value upto tens of TeVs. Nevertheless the Higgs scalars and the third family of squarks are presumed to have dispersed mass values at $ M_{GUT} $. In this computation the mass of third generation of squarks and Higgs scalars are zero. A vanishing $ A_{0} $ in our analysis is conjectured. {\color{blue}\cite{utpal}}. Computations obtained with the non universal scalar masses at  
\begin{center}
\begin{figure*}[htbp]
\centering{
\begin{subfigure}[]{\includegraphics[height=5.9cm,width=7.9cm]{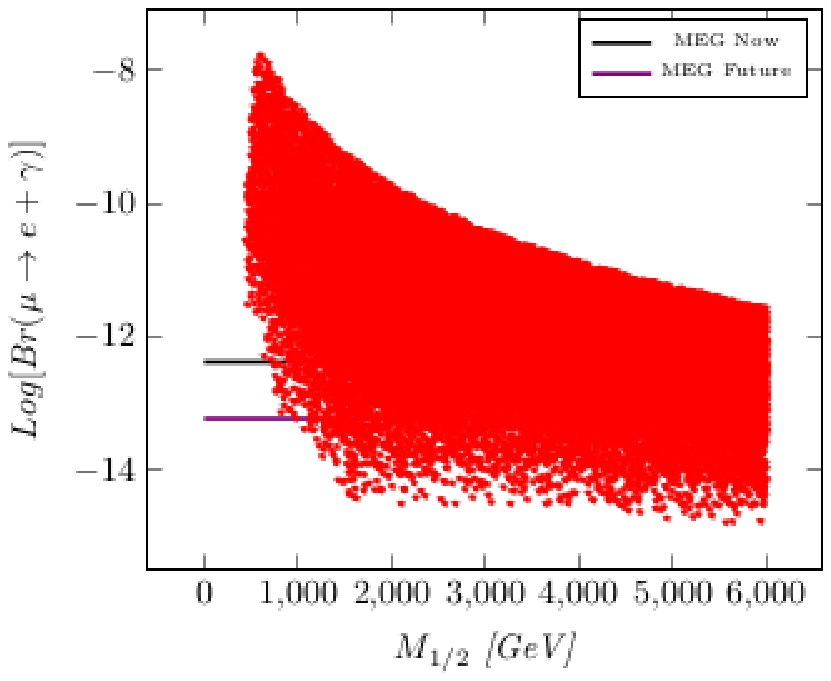}}\end{subfigure}
\begin{subfigure}[]{\includegraphics[height=6cm,width=8cm]{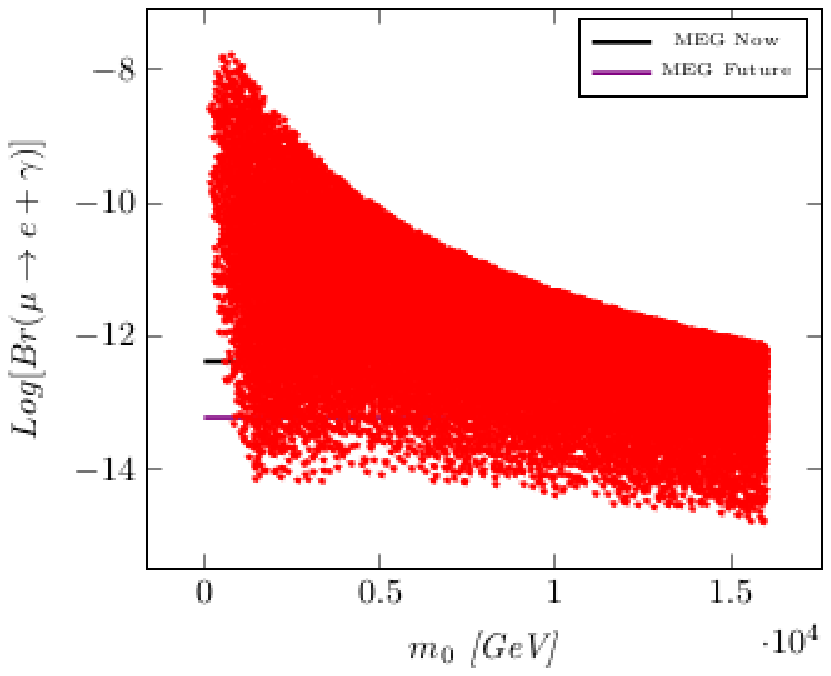}}\end{subfigure}\\
\begin{subfigure}[]{\includegraphics[height=5.9cm,width=7.9cm]{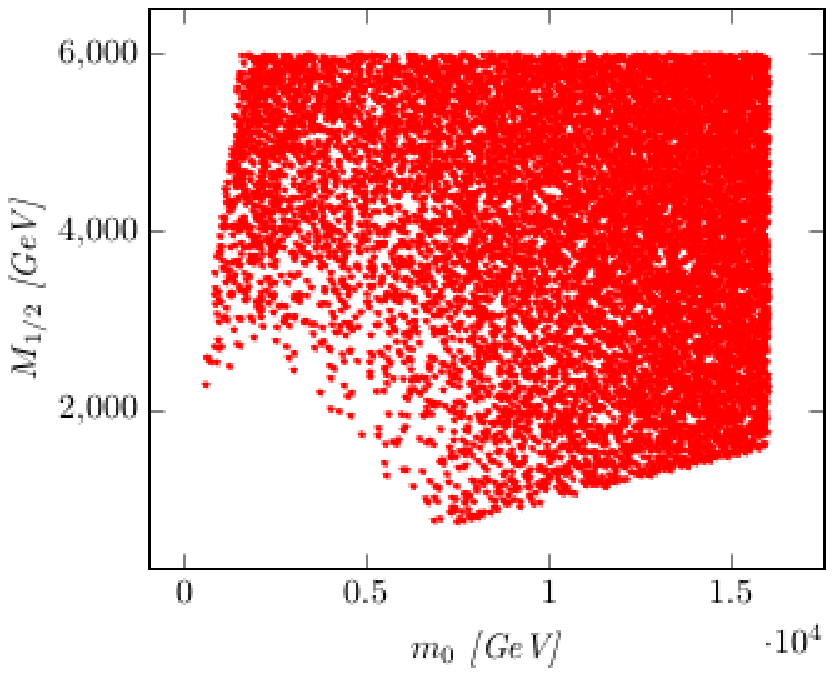}}\end{subfigure}
\begin{subfigure}[]{\includegraphics[height=6cm,width=8cm]{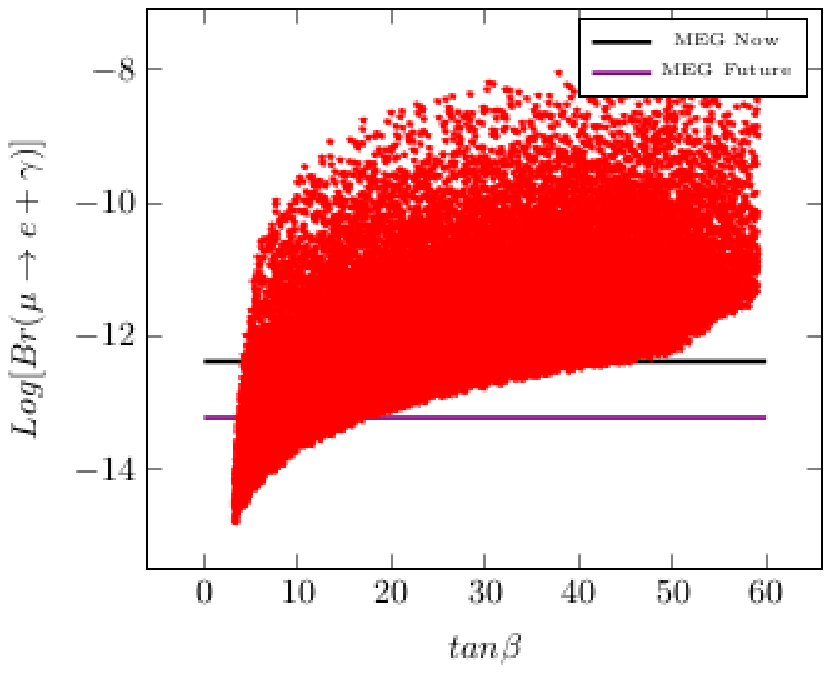}}\end{subfigure}\\

\caption{The results of the analysis are presented for NUSM case. In fig {\color{blue}7a, 7b}, different horizontal lines depicts the present (MEG 2016) and future MEG bounds for BR($ \mu $ $ \rightarrow $ e
 + $ \gamma $). Figs. {\color{blue}7c, 7d} shows the allowed space for different parameters, that is allowed by MEG 2016 bound.}}
\label{fig:2}
\end{figure*}
\end{center}
$ M_{GUT} $ is presented. In fig.7a, the soft SUSY parameter space as allowed by present and future MEG bounds on $BR(\mu \rightarrow e \gamma)$ is depicted. For the present MEG bound on Log$BR(\mu \rightarrow e \gamma)$  i,e -12.376, a large part of $ M_{1/2} $ parameter space survives. $M_{1/2}$ lies between  $ 1 TeV \hspace{.1cm}\text{TeV} \leq M_{1/2}\leq 6 \hspace{.1cm} \text{TeV}$. In fig.7b $ 2 \hspace{.1cm}\text{TeV} \leq m_{0}\leq 16 \hspace{.1cm} \text{TeV}$ exists for MEG constraint i.e -12.376. The fig.7c in the right panel shows $ m_{0} $ GeV Vs $ M_{1/2} $ GeV as constrained by MEG 2016 bound on BR$(\mu \rightarrow e \gamma)$. From fig. 7d it is seen that present MEG bound allows tan$ \beta $ to lie between 5 to 47, whereas future MEG bound restricts high values of tan $ \beta $ and limits itself to low values, where $ 5 \hspace{.1cm} \leq tan \beta \leq 18 \hspace{.1cm}$. It is found from fig. 8a that for Higgs boson mass around 125 GeV, $ m_{0} \geq $ 10 TeV is mostly allowed. For $ m_{h} \simeq $ 125.9 GeV the parameter space $ 10000 GeV \leq m_{0} \leq 16000 GeV $ is mostly recommended. From fig. 8b it is observed that for higgs mass range around $125\pm 2 $ GeV, $ M_{1/2} $ lies between 1.5 TeV $ \leq M_{1/2} \leq $ 5 TeV. From fig.8c, it is grasped Higgs mass of 125 GeV disfavours values of both low tan$ \beta \leq 15 $ and high tan$ \beta \geq 30 $. Fig. 8d and fig.8e represents favoured values of soft susy space detectable at future run of HE LHC.
\begin{center}
\begin{figure*}[htbp]
\centering{
\begin{subfigure}[]{\includegraphics[height=5.9cm,width=7.9cm]{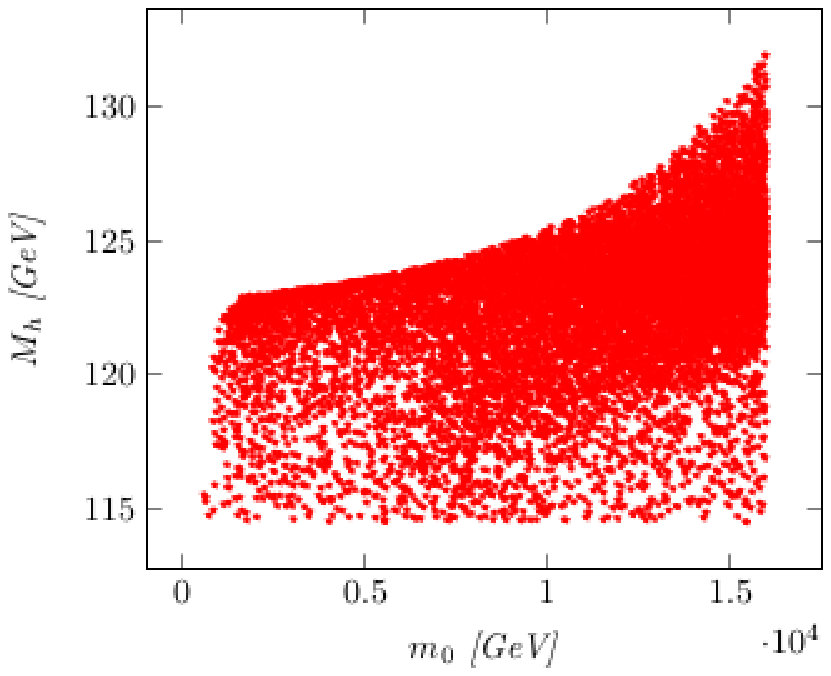}}\end{subfigure}
\begin{subfigure}[]{\includegraphics[height=6cm,width=8cm]{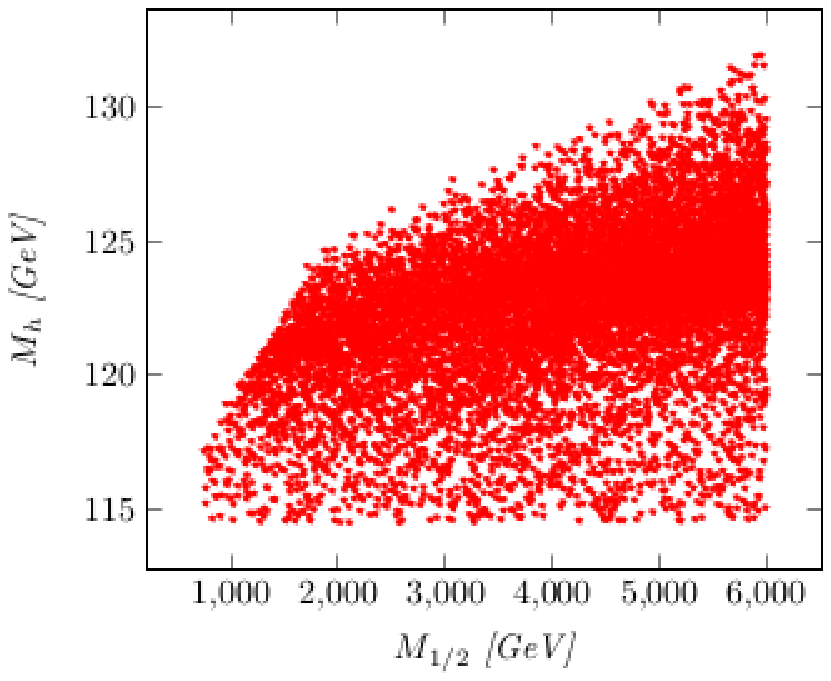}}\end{subfigure}\\
\begin{subfigure}[]{\includegraphics[height=5.9cm,width=7.9cm]{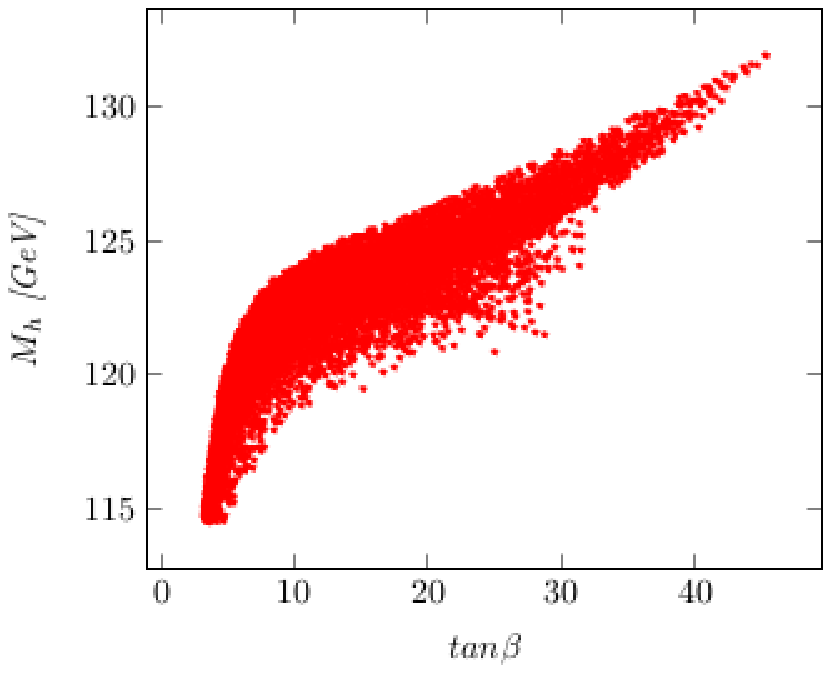}}\end{subfigure}
\begin{subfigure}[]{\includegraphics[height=6cm,width=8cm]{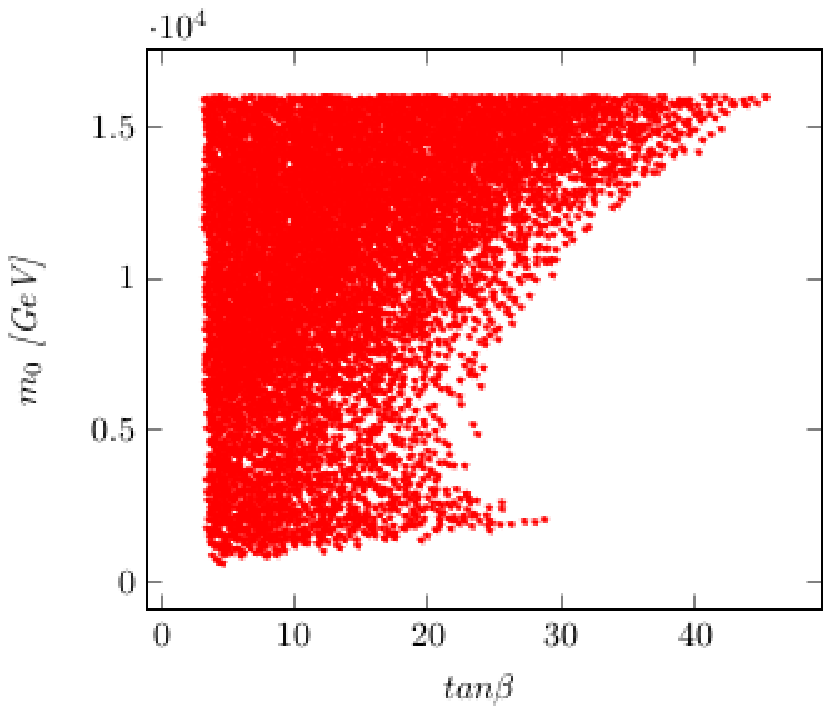}}\end{subfigure}\\
\begin{subfigure}[]{\includegraphics[height=6cm,width=8cm]{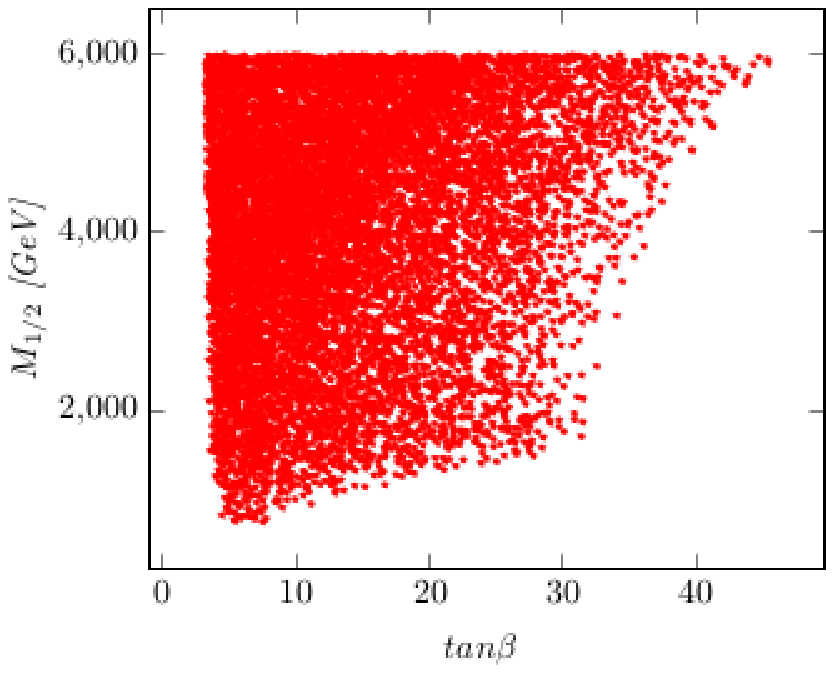}}\end{subfigure}\\
\caption{The results of the  computations presented for NUSM case. Fig {\color{blue}8a, 8b, 8c, 8d, 8e}, depicts the permitted space for different parameters, that is allowed by MEG 2016 bound by MEG collaboration.}}
\label{fig:2}
\end{figure*}
\end{center}
 In Tables II and III the relative research of the comparative study between different models is presented here. The new consequences in NUSM are the following:
\newline
1. Fragile $m_{0}$ susy space is allowed as contrasted to CMSSM.
\newline
2. A wider SUSY parameter space is favoured as compared to CMSSM and NUHM model.
\newline
3.For Higgs mass around 125 GeV, favoured values of tan $ \beta  $ are $ 12 \leq tan\beta \leq 28$ are allowed. Tan $ \beta $ values less than 30 are allowed.
\newline
4.The expected sensitivity of the MEG-II experiment which  is $6 \times 10^{-14}$ for three years of data taking restricts values of tan$ \beta $ to be less than 20 and these predict or forecast small LFV rates which is easily accessible from the figures.
\begin{table*}
\caption{Spartcle masses in this table are comparison between model CMSSM and NUHM.}
\label{}
\begin{tabular*}{\textwidth}{@{\extracolsep{\fill}}lrrrrl@{}}
\hline\noalign{\smallskip}
\hline
Range of parameters allowed by & Range of parameters allowed by
\\  for BR $\left( \mu \rightarrow e \gamma\right) < 4.2\times 10^{-13}$  & BR $\left(\mu \rightarrow e\gamma\right) < 4.2\times 10^{-13}$  
\\  MEG 2016, for CMSSM & MEG 2016, for NUHM  \\
 \hline
\textbf{1}.Fig 3a:& \textbf{1}.Fig 5b: For  MEG 2016, $M_{1/2}\geq$ 1 TeV 
\\$M_{1/2}\geq$ 3.2 TeV by MEG 2016 & Fig: 5a, $m_{0} \geq$ 2.2 TeV  
 \\ Fig 3b: $m_{0} \geq$ 4.2 TeV & for BR$\left( \mu \rightarrow e \gamma\right)<10^{-13}$. 
  \\ 
\textbf{2}.Fig 4b (MEG 2016), : & \textbf{2}.Fig. 5c  for MEG 2016 of BR $\left(\mu \rightarrow e\gamma\right)$
\\ tan $ \beta $ lies in the range, 2 $ \leq $tan$ \beta  \leq$ 10 & tan $ \beta $ lies in the range, 3 $ \leq $tan$ \beta  \leq$60 
\\  for BR $\left( \mu \rightarrow e \gamma\right) < 4.2\times 10^{-13}$  & for BR $\left( \mu \rightarrow e \gamma\right) < 4.2\times 10^{-13}$   \\
\textbf{3}.Fig 4g: $ A_{0} $ lies in the range,  & \textbf{3}. Fig: 6b, $ A_{0} $ lies in the range, \\ $ A_{0}\sim $ -9000 GeV and$ A_{0}\sim $ 12000 GeV    & -10000 GeV $\leq A_{0} \leq $ -1000 GeV, \\ for $m_{h} = 125.9 $ GeV 
 & for $ m_{h} = 125.9\hspace{.1cm} \text{GeV} $
\\ 

\hline
\end{tabular*}
\end{table*}

\begin{table*}
\caption{Spartcle masses in this table are comparison between model NUSM and NUHM.}
\label{}
\begin{tabular*}{\textwidth}{@{\extracolsep{\fill}}lrrrrl@{}}
\hline\noalign{\smallskip}
\hline
Range of parameters allowed by & Range of parameters allowed by
\\  for BR $\left( \mu \rightarrow e \gamma\right) < 4.2\times 10^{-13}$  & BR $\left(\mu \rightarrow e\gamma\right) < 4.2\times 10^{-13}$  
\\  MEG 2016, for CMSSM & MEG 2016, for NUHM  \\
 \hline
\textbf{1}.Fig 7a:& \textbf{1}.Fig 5b: For  MEG 2016, $M_{1/2}\geq$ 1 TeV 
\\$M_{1/2}\geq$ 1 TeV by MEG 2016 & Fig: 5a, $m_{0} \geq$ 2.2 TeV  
 \\ Fig 7b: $m_{0} \geq$ 2.2 TeV & for BR$\left( \mu \rightarrow e \gamma\right)<10^{-13}$. 
  \\ 
\textbf{2}.Fig 7b (MEG 2016), : & \textbf{2}.Fig. 5c  for MEG 2016 of BR $\left(\mu \rightarrow e\gamma\right)$
\\ tan $ \beta $ lies in the range, 5 $ \leq $tan$ \beta  \leq$ 50 & tan $ \beta $ lies in the range, 3 $ \leq $tan$ \beta  \leq$60 
\\  for BR $\left( \mu \rightarrow e \gamma\right) < 4.2\times 10^{-13}$  & for BR $\left( \mu \rightarrow e \gamma\right) < 4.2\times 10^{-13}$   \\
\textbf{3}.Fig 8c: $ tan\beta $ lies in the range,  & \textbf{3}. Fig: 6d, $ tan\beta$ lies in the range, \\ 15 $ \leq $tan$ \beta  \leq$ 30    & 8 $ \leq $tan$ \beta  \leq$ 13\\ for $m_{h} = 125.9 $ GeV 
 & for $ m_{h} = 125.9\hspace{.1cm} \text{GeV} $
\\ 

\hline
\end{tabular*}
\end{table*}
\section{Conclusion}
The see-saw mechanism is very encouraging in the sense that an intermediate mass scale exists for Majorana neutrinos which solves the mystery of the tiny active neutrino masses neverthelss also describing the absence of right-handed
neutrino effects in low energy laboratory scale. The see-saw mechanism is not an impelling extension of the SM since the Higgs boson mass would likely blasts to the see-saw scale owing to its quadratic divergent radiative corrections to the Higgs scalar masses. Supersymmetry stabilizes the hierarchy problem and steadfast
the Higgs mass so the weak scale can mutually stays along with the Majorana mass scale (and the GUT and Planck scales). The aim of this paper is to present forecast for LFV
process $ \mu \rightarrow e + \gamma $ within credible SUSY models that are consistent with LHC Run 2 results. Within SUSY models, LFV mechanisms should happen, possibly at an observable level. The motivation for this paper is to present predictions for cLFV
processes within various SUSY models that are happening with LHC Run 2 results. These should naturally include SUSY models with radiatively-driven naturalness [22] which requires a 125 GeV Higgs mass along with multi-TeV soft terms (as implied by LHC data) nevertheless at the same time it avoids the fine-tunings attached with the Little Hierarchy problem. Here, the current and projected reaches of LFV search $ \mu \rightarrow e + \gamma $ in MEG II and MEG collaboration within CMSSM, NUHM and NUSM SUSY model where neutrinos are generated with a type-II seesaw mechanism is examined. The soft SUSY parameter space is constrained within the HE-LHC reach. The NUHM model is well inspired in that it admits for weak scale naturalness accompanying with a 125 GeV Higgs mass and sparticles beyond LHC Run 2 limits. 
\par 
The value of Higgs mass as measured at LHC, latest global data on the reactor mixing angle $ \theta_{13} $ for neutrinos, and latest constraints on BR($ \mu  \rightarrow e  \gamma $) as projected by MEG collaboration{\color{blue}\cite{Adam}},{\color{blue}\cite{Baldini}} is used here. In CMSSM a very heavy $ M_{1/2} $ region is allowed by projected sensitivity of MEG II experiments which is 6$ \times 10^{-14} $, nevertheless in NUHM case comparatively a low $ M_{1/2} $ region is also favoured. Further the non universal scalar mass model (NUSM) is also studied. As compared to CMSSM, in NUHM, a broader soft susy space is allowed. It is found that in NUSM, a wide spectra of susy parameter space is admitted, as compared to both CMSSM and NUHM. In NUHM, it is seen that favoured values of $ \vert A_{0}\vert $ consistent with the newly discovered Higgs boson mass are negative and are shifted towards lighter side (compared to CMSSM). It is perceived that in NUSM as Log[BR($ \mu  \rightarrow e  \gamma $)]  decreases from present MEG bound to around to future MEG II projected sensitivity, a wide region of soft susy parameter space of $ M_{1/2} $ from 1 TeV to almost 6 TeV is allowed. The computations in this paper show that for natural SUSY with the model CMSSM, rates
for BR($ \mu  \rightarrow e  \gamma $) are already ruled out by the projected MEG II sensitivity experiment for $ S_{4} \times Z_{n} $ flavor symmetric SUSY SO(10) theory 
scenarios. The non universal boundary conditions in NUSM can probe experimental indications for the production of supersymmetric particles and can entertain detector set up to ensure that various supersymmetric models can probe signatures at HE LHC.  Observation of sparticles at HE/HL LHC, could help us to distinguish among CMSSM, NUSM and NUHM models, in context to the limits put by cLFV decays. This in turn leads to a motivation for examining various SUSY theories beyond standard model.
\begin{acknowledgements}
This work was supported in part by the Department of Physics, Office of High Energy Physics, IISC, Bangalore, India. The computing for this paper is performed by SuSeFLAV software. 
\end{acknowledgements}

\end{document}